\documentstyle[11pt,aaspp4,flushrt]{article}
\def\Rin{R_{\rm in}}

\def\Rout{R_{\rm out}}

\def\Rg{R_{\rm g}}
\def\Ka{{\rm K}{\alpha}}
\def\etal{et al.\ }

\def\ginga{{\it Ginga}\ }

\def\MEdd{M_{\rm Edd}}
\def\chir{\chi^2_{\nu}}
\def\dof{\,{\rm d.o.f.}}
\def\Firr{F_{\rm irr}}
\def\feab{[{\rm Fe}]}
\def\LX{L_{\rm X}}
\def\mdot{{\dot m}}

\def\hrec{h_{\rm rec}}

\righthead{Relativistically smeared X-ray reprocessed spectra of GS~2023+338}
\lefthead{P.T. \.{Z}ycki, C. Done and D.A. Smith}

\begin{document}

\title {Relativistically smeared X-ray reprocessed components in the 
GINGA spectra of GS~2023+338}

\author{Piotr T. \.{Z}ycki and Chris Done}
\affil{Department of Physics, University of Durham, South Road, 
Durham DH1 3LE, England; piotr.zycki@durham.ac.uk, chris.done@durham.ac.uk}

\and

\author{David A. Smith}
\affil{Department of Physics and Astronomy, University of Leicester, 
       University Road, Leicester LE1 7RH, England; das@star.le.ac.uk}

\begin{abstract}

We present results of spectral analysis of \ginga data obtained
during the decline phase after the 1989 outburst of GS~2023+338 (V404 Cyg).
Our analysis includes detailed modelling of the effects of X-ray
reflection/reprocessing. We have found that (1) the contribution of
the reprocessed component (both continuum and line) corresponds to 
the solid angle of the reprocessor as seen from the X-ray source of 
$\Omega\approx (0.4-0.5)\times 2\pi$, (2) 
the reprocessed component (both line and continuum) is broadened (``smeared'')
by kinematic and relativistic effects, as expected from the accretion disk
reflection. We discuss the constraints these results give on 
various possible system geometries.

\end{abstract}

\keywords{accretion, accretion disk -- black hole physics -- 
binaries: close --  stars: individual (V404 Cyg) -- X-ray: stars}

\section{Introduction}

Some of the strongest evidence for the existence of accretion disks
around black holes has come from X-ray observations of the
relativistically smeared iron $\Ka$ line profile in Active
Galactic Nuclei (\markcite{ta95}Tanaka \etal 1995;
\markcite{iw96} Iwasawa \etal 1996;
\markcite{na97} Nandra \etal 1997). 
This fluorescence line is produced by hard X--ray illumination
of the accreting material, and the combination of high orbital
velocities and strong gravity in the vicinity of a black hole gives
the line a characteristically skewed, broad profile 
(\markcite{fa89}Fabian \etal 1989; \markcite{la91} Laor \etal 1991). 
A reflected continuum should also accompany
this line (e.g.\ \markcite{lw88} Lightman \& White 1988; 
\markcite{gf91} George \& Fabian 1991; \markcite{ma91} Matt,
Perola \& Piro 1991), and the amplitude of both reprocessed components
gives constraints on the solid angle subtended by the accreting
material, its inclination, elemental abundance and ionization
state. Both the amount of reflection/fluorescence and the intensity of
the relativistic effects on the observed line profile strongly support
the idea that the accretion disk extends down to the last stable orbit
in AGN.
The Black Hole Candidates (BHC) show many X--ray spectral
similarities to AGN, plausibly because both involve the same physical
processes of disk accretion onto a black hole 
(\markcite{in93}Inoue 1993; 
\markcite{ue94} Ueda, Ebisawa \& Done 1994;
\markcite{gi97} Gierli\'{n}ski \etal 1997).

Many Soft X-ray Transients (SXT) are known to be BHC.  Their mass
accretion rate varies over at least 2 orders of magnitude from
outburst (where it is at about Eddington) through the decline on
time-scales of months (see \markcite{ts96} Tanaka \& Shibazaki 1996 for recent
review).  A number of these systems were observed by \ginga and the
spectra obtained are amongst the best in which to investigate the
overall effects of X--ray reprocessing.
Moderate spectral resolution (18\% at 6 keV) is more than
compensated by very high signal--to--noise, broad band pass (from 1
keV up to 20-30 keV) and our relatively good understanding of the
instrument.

In this Letter we present results of spectral analysis of data
obtained during the 1989 outburst of GS~2023+338 (V404 Cyg; 
\markcite{ki89} Kitamoto
\etal 1989).  The outburst was well covered by \ginga from its initial 
detection by All Sky Monitor on May 22 until November 1989
(\markcite{tl95}Tanaka \& Lewin 1995).  We have selected two data sets 
where there is little short time-scale spectral variability,
obtained June 20 and July 19-20, respectively.
We analyse these using models of X--ray
reprocessing that consistently connect the properties of the iron
$\Ka$ line with the properties of the reflected continuum. We show
that the reflected component is present in the spectra, that it is smeared
by kinematic and relativistic effects as expected from an accretion disk,
and that both its normalization and amount of smearing imply
that the disk does not extend down to the last stable orbit.

A full analysis of the data covering the outburst and entire decline phase
will be presented in future paper (\.{Z}ycki et al., in preparation, hereafter
paper II).

\section{Data selection and reduction}

The data were extracted from the original FRF's 
using the \ginga reduction software at Leicester University.
Paradoxically, background subtraction poses a problem for a source
as bright as GS~2023+338 as the background monitors are
strongly contaminated by source counts. However, the background can be
estimated from nearby observations at similar points in the satellite orbit,
and its fractional contribution is low ($\le 3$\% below 10 keV).
Full details of our method of background subtraction will be given in 
paper II.
We allow for $0.5\%$ systematic error in the data.

\section{Model}

The model components are: multi-temperature accretion disk spectrum
(``disk black body''; \markcite{mi84} Mitsuda \etal 1984)
and a power law with its Compton reflection from a possibly ionized 
medium including the iron $\Ka$ line emission. 
The Compton reflected component is computed using the XSPEC version 9.01 model
``pexriv'' (Magdziarz \& Zdziarski\markcite{mz95} 1995) with the 
ionization parameterised by the ionization parameter,
$\xi\equiv \LX/n r^2$ as in Done \etal\markcite{do92} (1992).
We assume elemental abundances of Morrison \& McCammon \markcite{mm83} 
(1983) with 
the exception of the iron abundance, $\feab$, which is a free
parameter. The iron $\Ka$ line is computed using modified Monte Carlo 
simulations code of \.{Z}ycki \& Czerny \markcite{zc94} (1994). We updated the
Fe $\Ka$ line energies, fluorescent yields and Fe K-shell edge energies
after Kaastra \& Mewe \markcite{km93} (1993).

The reprocessed component can then be ``smeared'' to simulate the
relativistic and kinematic effects of disk emission
(see e.g.\ \markcite{fa89} Fabian \etal 1989; \markcite{ro96} 
Ross, Fabian \& Brandt 1996). We assume a non-rotating black hole so
the model is parameterised by the
inner and outer radius of the disk, $\Rin$ and $\Rout$ respectively,
and the radial distribution of irradiation emissivity. 
We fix the form of the emissivity,
$\Firr(r)\propto r^{-3}$ as expected from coronal illumination,
fix $\Rout$ at $10^4\,\Rg$, since it is an appropriate value for
LMXRB ($\Rg\equiv GM/c^2$) and fit $\Rin$ and the normalization of
the reprocessed component, $f\equiv\Omega/2\pi$, where $\Omega$ is
the solid angle of the reprocessor as seen from the X-ray source.
We emphasize that $\Rin$, $\Rout$ and $\Firr(r)$ describe only 
the smearing effect, not $\Omega$.

We assume that the inclination of the system is $i=56^{\circ}$ 
(\markcite{pa96}Pavlenko \etal 1996).

\section{Results of model fitting}

We begin with the simplest possibility, that is a power law spectrum with
a narrow line at 6.4 keV. We than add the reflected continuum assuming first
that it is un-ionized, non-smeared
and the iron abundance is 1 relative to cosmic value.
In the next step we fit $\xi$ and $\feab$ and finally we introduce 
the effect of smearing. This multi-step procedure is important (1)
to demonstrate the significance of reprocessing and (2) since
the moderate spectral resolution of \ginga
means that the broadening of iron spectral features due to ionization
can mimic the relativistic smearing. 

For the June 20th data set, the 
simplest model (model 0 in Table~1) gives $\chir=1.9$ 
and it is thus not acceptable. Adding the reprocessed component is
highly significant even in the simplest version (model A),
$\chi^2=19.9/25\dof$ ($\chir=0.80$).
We note that the normalization of the reflected component is
significantly smaller than 1, $f=0.46\pm 0.04$.

The fit can be improved by allowing for ionized reflection (model B). 
Assuming
the reprocessor temperature $T=10^6\,$K (for the purpose of computing
ionization balance only), the best fit has $\chi^2=15.2/24\dof$
for $\xi=0.1^{+4}$. If, instead, we let iron abundance be free whilst
fixing $\xi=0$ (model C), we obtain $\chi^2=19.1/24\dof$ for 
$\feab=1.20\pm 0.35$, i.e.\ $\feab$ is not a significant parameter.
The best fit with both $\xi$ and $\feab$ free (model D)
has $\chi^2=14.9/23\dof$ for $\xi=0.1$ but again $\feab$ is consistent
with the cosmic value.

\begin{figure}
 \plottwo{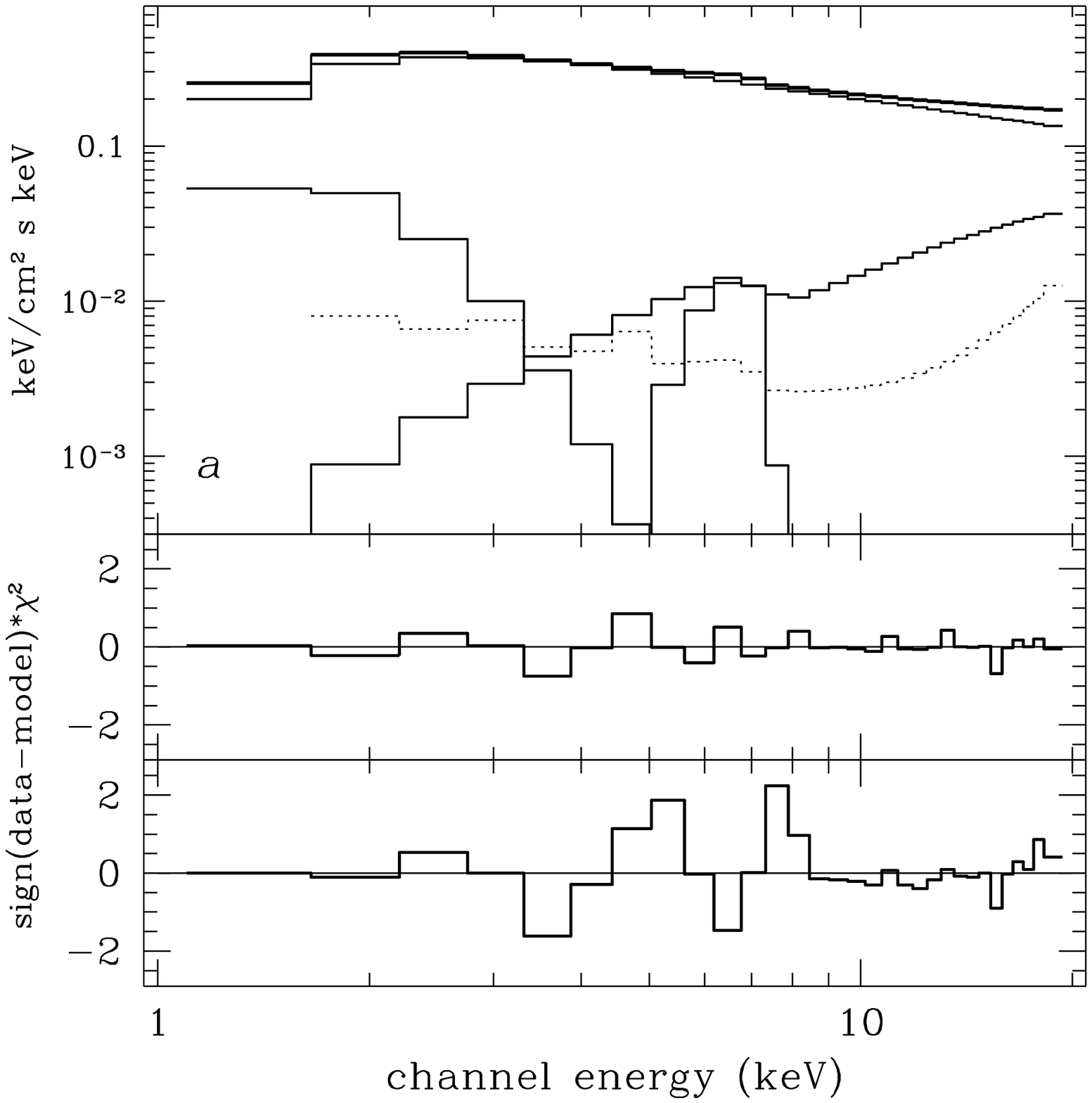}{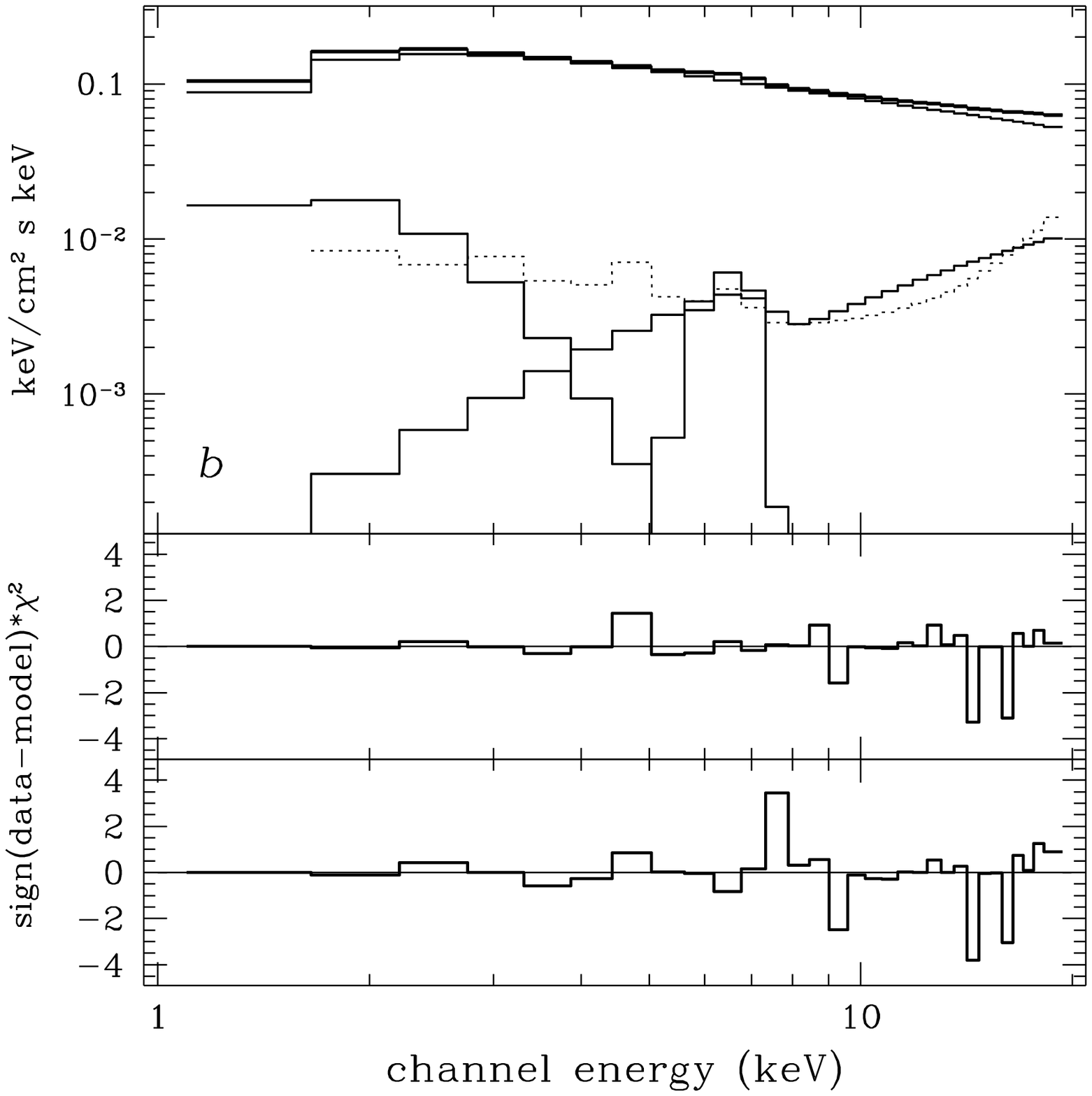}
 \figcaption{Best fit spectra (models E in Tables~1.\ and 2.; upper panels), 
  their $\chi^2$ residuals (middle panels) and the $\chi^2$ residuals for 
  models D i.e.\ without kinematic smearing.  The dotted histograms 
  show the background.  Left panels {\it (a)}\/ are for June  data, 
  right panels {\it (b)}\/ for July data.
 \label{fig:june}}
\end{figure}

We now introduce the effect of smearing (model E; 
both $\xi$ and $\feab$ are left free as well). 
This results in a decrease of $\chi^2$ by $\Delta\chi^2=8.9$
which is significant at more than 99.9\% confidence level 
(the F-test for one additional parameter). 
The best fit inner radius is $\Rin=25^{+45}_{-11}\,\Rg$. Figure~\ref{fig:june}
shows the spectrum and residuals of the best fit models D and E.

A similar progression in quality of the fit is given by the July spectrum
(Table~2; Figure~1), where again the relativistic smearing effects are
significantly present in the data.

We have also tried a model with $\xi$ changing as a function of radius,
$\xi(r)\propto r^{\alpha}$, but with no relativistic smearing. 
Best fit of the model for the June data has
$\chi^2=14.6/22\dof$, while for the July data $\chi^2=20.1/22\dof$

\section{Discussion}

\subsection{Primary continuum}

\label{sec:prim}

From fits we have $\Gamma\sim 1.7$, while contemporaneous high energy data 
show
a cutoff at about 100 keV (\markcite{su91}Sunyaev \etal 1991).  
The source must
then be marginally optically thin if the primary X-rays are produced by 
thermal
Comptonization, with $\tau_{\rm T}\sim 0.5$ and $1.5$, in a disk or 
sphere geometry, respectively (Titarchuk\markcite{ti94} 1994).

\subsection{Geometry}

\label{sec:geom}

The results of spectral modelling clearly show that the reprocessed component 
is present in these data.
The required smearing cannot be explained by a radial distribution of 
ionization since the data strongly favour a low and uniform ionization.
Thus the relativistic effects in an accretion disk/black hole system
are the most plausible explanation. The reflected fraction for both
continuum and line is roughly $\sim 0.5\times$ that expected from an
isotropically illuminated flat disk. Thus the line is {\em not}\/ depleted by
Auger ionization masked by relativistic smearing (Ross \etal 1996). The line 
is
weak because the covering fraction of the reflecting material is small, and
Auger ionization is ruled out by the low ionization state of the disk.  The 
low
covering fraction does not seem to be an artifact of super--solar abundances: 
fixing $f=1$
and allowing the abundance to be free results in a poorer fit, with 
$\chi^2=19/23\dof$ and $16/23\dof$ for the June and July data,
 respectively, and the overall abundances, $A\sim 8$.  It is also
commonly seen in the persistent BHC: Cyg~X--1 and GX~339-4 (Done \etal 1992;
Gierli\'{n}ski \etal 1997; Ueda \etal 1994),  so is {\em not}\/ some 
time dependent effect of disk evolution in transient systems, but rather 
represents a significant geometrical constraint.  Another constraint 
comes from the fact that
the observed relativistic smearing is less than that expected from a central
point source illuminating a flat disk which extends down to $6\Rg$.  

One possible geometry is a spherical or flattened X--ray source centered on 
the
black hole as originally proposed by Thorne \& Price\markcite{tp75} (1975) 
for Cyg X--1. 
A spherical source with $\tau_{\rm T}\gg 1$ gives $f=0.5$ naturally 
(Done \etal 1992),
as the X--ray emission only escapes into a hemisphere tangential to the source
surface. However, here the source is probably marginally optically thin with
$\tau_T\sim 0.5-2$ (see \S~\ref{sec:prim}). 
This can still give a reduction in $f$
since there is a hole in the inner disk at $r\le \Rin$. Some of the photons
escape without illuminating the disk, and the remaining reflection fraction is
diluted by hard X--ray photons produced on the other side of the disk. The
relativistic smearing constraints then imply that the the source emissivity is
less steep than $\propto r^{-3}$ and/or the inner disk disk truncates at
$\Rin> 6\,\Rg$ and/or the outer disk flares.  
There are few real physical constraints on the source emissivity. 
Local release of gravitational energy in a Keplerian disk 
gives a luminosity $\propto r^{-3}$,
but non--local mechanisms could also operate, perhaps even producing a 
constant source $\propto r^0$. Outside of the source the illumination 
is approximated by a central
point source i.e.\ $\propto r^{-3}$ but within the spherical source 
the illumination produces an emissivity of $\propto r^{-2}$ 
(for luminosity $\propto r^{-3}$). This
can reduce the most strongly smeared components to the observed level, though
truncation of the inner disk radius may also be required. 
The geometry might then
be in accord with recently proposed scenario of SXT evolution after outbursts 
(Esin, McClintock \& Narayan\markcite{es97} 1997 and references therein)
although the overall behavior of the source does not seem to follow 
the proposed scenario.

Small active regions on the disk (a ``patchy'' corona), 
perhaps powered by magnetic reconnection
give another possible geometry (Haardt, Maraschi \& Ghisellini
\markcite{hmg94} 1994, Stern \etal\markcite{st95} 1995). 
Physical support for such a picture may come from recent models of
accretion disk viscosity as an MHD disk dynamo (Hawley, Gammie \& Balbus 
\markcite{hgb96} 1996). However, these models give $f\sim 1$ unless the 
height of the
reconnecting regions, $\hrec$, is large compared to the size of the disk.  
This
would require that the reconnecting regions be concentrated along the inner 
disk radius $\Rin$, and that $\hrec\sim \Rin$, since the outer disk is
expected to be very large. This would also go some way to satisfying the 
smearing constraints, as the irradiation would then be constant 
from $\sim \Rin$ to $\sim \Rin+\hrec$ rather than $\propto r^{-3}$.

Another geometry that has been proposed, though with rather less physical
motivation, is a continuous corona overlying the accretion disk (Haardt \&
Maraschi\markcite{hm93} 1993). A fraction 
$\sim 1- \exp (-\tau/\cos i)\sim 0.5$ of the reflected
spectrum is itself Compton scattered by the X--ray emitting corona, and loses
its characteristic spectral shape (Haardt \etal\markcite{ha93} 1993). 
The energy generation 
would be  expected to be $\propto r^{-3}$, so the smearing constraints
are a problem unless the inner disk truncates, or the outer disk flares.
%One argument for these models is that the 2 -- 20 keV spectral index is
%an excellent match to that predicted (Haardt \etal\markcite{ha93}1993), 
%showing that the
%continuum is consistent with Compton cooling of the hot electrons on seed
%photons with luminosity $\sim 1-2\times$ that of the hard X--ray
%source. 
However, this simple scenario can be ruled out from detailed spectral
fitting of Cyg X--1 (Gierli\'{n}ski \etal\markcite{gi97}1997;
Poutanen, Krolik \& Ryde\markcite{pkr97} 1997), 
although perhaps this merely
points to added complexity such as temperature/optical depth structure in the
corona.

Additional independent constraints on the geometry are provided by 
low normalization 
of the soft component and weak ionization of the reflecting medium. 
They both support the idea of the disk being truncated at $\sim 30\,\Rg$
as then both the expected temperature of Shakura--Sunyaev\markcite{ss73} 
(1973) disk
emission component (for $\mdot= 0.01\,\MEdd$ and $M=10\,M_{\sun}$),
$T\approx 0.15\,$keV and the ionization parameter,
$\xi\sim 1$, would be in agreement with our results. However, low $\xi$
may also mean that the disk is much denser than the SS solution due to
e.g.\ coronal dissipation of energy 
(e.g.\ Svensson \& Zdziarski\markcite{sz94} 1994). 

\section{Conclusions}

We have performed spectral analysis of \ginga data of soft X-ray 
transient GS~2023+338 (V404~Cyg)
obtained one and two months after its 1989 outburst. We have found that
\begin{itemize}
 \item the Compton reflected continuum and iron fluorescent $\Ka$ line
are present in the spectrum,
 \item the properties of both reprocessed components (normalizations,
ionization parameters) are in agreement,
 \item the data require both reprocessed components to be broadened and 
smeared,
 \item the smearing is consistent with being due to reflection from 
a Keplerian disk,
 \item both the normalization of the reflected continuum and the amount
of smearing constrain the geometry. 
\end{itemize}

\acknowledgements

This research 
made use of data obtained from the Leicester Database and Archive 
Service at the Department of Physics and Astronomy, Leicester University.
CD acknowledges support from a PPARC Advanced Fellowship. Work of PTZ
was partly supported by grant No.\ 2P03D00410 of the Polish State
Committee for Scientific Research.

\clearpage

\clearpage

\begin{deluxetable}{ccccccccc}
 \footnotesize
 \tablewidth{0pt}
 \tablecaption{June 20th data \label{tab:june}}
\tablehead{
\colhead{model} & \colhead{$kT\,$keV} & \colhead{$\Gamma$} & \colhead{$\xi$} &
\colhead{$\feab$} & \colhead{$f$} & \colhead{$\Rin$ ($\Rg$)} & 
                                     EW (eV)   &   \colhead{$\chi^2/\dof$}
}
\startdata
 0  &  $1.51\pm 0.08$         & $1.34\pm 0.02$  & \nodata & \nodata & \nodata 
         & \nodata &        $90^{+23}_{-14}$ &   46.8/25                \nl
 A  &  $0.36\pm 0.04$         & $1.66\pm 0.01$         &  0(f)      &
          1(f)       & $0.46\pm 0.04$   & \nodata  & \nodata   & 19.9/25  \nl
 B  &  $0.36^{+0.09}_{-0.04}$ & $1.66^{+0.01}_{-0.03}$ & $0.1^{+4}$   &
          1(f)       & $0.45^{+0.04}_{-0.06}$ & \nodata & \nodata & 15.2/24\nl
 C  &  $0.38^{+0.17}_{-0.05}$ & $1.64\pm 0.04$         &  0(f)      &
  $1.2\pm 0.35$      & $0.43\pm 0.06$         & \nodata & \nodata & 19.1/24\nl
 D  &  $0.37^{+0.18}_{-0.05}$ & $1.65\pm 0.03$         & $0.1^{+4}$   &
  $1.0^{+0.4}_{-0.2}$ & $0.44\pm 0.06$        & \nodata & \nodata & 14.9/23\nl
 E  & $0.39^{+0.25}_{-0.05}$ & $1.64\pm 0.04$          & $0^{+5}$   & 
  $1.5^{+0.7}_{-0.5}$ & $0.47\pm 0.06$ & $25^{+45}_{-11}$ & \nodata &
                                                            $6.0/22$ \nl

\enddata
\end{deluxetable}

\clearpage

\begin{deluxetable}{ccccccccc}
 \footnotesize
 \tablewidth{0pt}
 \tablecaption{July data}
 \label{tab:july}
\tablehead{
\colhead{model} & \colhead{$kT\,$keV} & \colhead{$\Gamma$} & \colhead{$\xi$} &
\colhead{$\feab$\tablenotemark{a}} & \colhead{$f$} & 
\colhead{$\Rin$ ($\Rg$)} & EW (eV)  & \colhead{$\chi^2/\dof$}
}
\startdata
 0  &  $1.4 \pm 0.1$ & $1.45\pm 0.25$  & \nodata & \nodata & \nodata & 
    \nodata &                $93\pm 14$ &       57/25                   \nl
 A  &  $0.36\pm 0.03$         & $1.72\pm 0.015$        &  0(f)      &
          1(f)       & $0.40\pm 0.05$ & \nodata & \nodata & 34.6/25        \nl
 B  &  $0.44^{+0.40}_{-0.09}$ & $1.68\pm 0.04$       & $13^{+25}_{-12}$ &
          1(f)       & $0.32\pm 0.08$ & \nodata & \nodata & 22.9/24        \nl
 C  &  $0.42^{+0.30}_{-0.06}$ & $1.68\pm 0.03$         &  0(f)      &
  $1.7^{+0.8}_{-0.5}$ & $0.35\pm 0.05$   & \nodata & \nodata  & 26.9/24    \nl
 D  &  $0.42^{+0.50}_{-0.06}$ & $1.68^{+0.04}_{-0.06}$ & $0.1^{+40}$ &
  $1.5\pm 0.6$ & $0.35^{+0.05}_{-0.10}$ & \nodata & \nodata  & 21.5/23   \nl
 E   & $0.47^{+0.35}_{-0.09}$  & $1.67^{+0.04}_{-0.05}$ & $0^{+20}$  &
     $2^{+1}_{-0.8}$ & $0.37^{+0.05}_{-0.08}$ & $35^{+110}_{-21}$ &
                                                  \nodata &  $15.4/22$  \nl

\enddata

\tablenotetext{a}{Upper limit on $\feab$ is 3}
\end{deluxetable}

\end{document}